# A Multilevel Network-assisted Congestion Feedback Mechanism for Network Congestion Control

Inayat Ali *, Seungwoo Hong and Tae Yeon Kim

*Electronics and Telecommunications Research Institute (ETRI), Daejeon, South Korea*

ABSTRACT

Network-assisted congestion control leveraging Explicit Congestion Notification (ECN) is an effective way to deal with congestion issues on the Internet. However, we believe that the existing ECN mechanism in the TCP/IP protocol stack may require further optimization to effectively address the evolving congestion challenges introduced by emerging technologies like immersive AR/VR applications and the burgeoning field of the Internet of Things (IoT). To that end, we propose a multilevel congestion notification mechanism called Enhanced ECN (EECN) that leverages the existing two ECN bits in the IP header to notify two levels of congestion in the network and uses the corresponding two bits in the TCP header to negotiate EECN during the handshake and echo congestion experienced back to the sender. Additionally, we propose a congestion control mechanism that triggers different congestion control responses based on the average RTT and multilevel congestion feedback received from the network, which yields promising results, highlighting the effectiveness of utilizing multilevel congestion feedback. The proposed EECN mechanism reduces packet drop by 70% compared to ECN, by 95% compared to TCP New Reno without ECN, and by 40% compared to VCP. The packets marked are reduced by 96% compared to ECN and 76% compared to VCP. Furthermore, the proposed approach reduces flow completion time by 61% compared to ECN and enhances the throughput of short-lived network flows, which are particularly pronounced in IoT environments.

## 1. Introduction

[1] Congestion control is not a new issue and has existed since the early stages of the Internet's evolution. However, it persists as a relevant challenge due to the multitude of new immersive Internet applications and their latency and bandwidth requirements. Challenges caused by immersive technologies may result in the need for new network architectures, transport protocols, and congestion control mechanisms. Furthermore, short-lived network flows account for most of the Internet traffic but have been overlooked by standard transport protocols and Active Queue Management (AQM) approaches [1]. Approximately 99% of Internet flows carry data less than 100 KB [2], requiring only a few Round Trip Times (RTTs) to complete, typically concluding before the TCP slow-start phase concludes and beginning to transmit at an optimal congestion window (cWnd) [3, 1]. The significance of short-lived network flows is particularly prevalent in Internet of Things (IoT) environments due to the frequent, small data transmissions characteristic of IoT devices [4].

Two main approaches are employed to address congestion issues in the network. The first category of protocols adopts end-host-based methods, detecting congestion through packet drops or using complex measurements at the end hosts to control flow [5–8]. However, protocols in this category often face resource underutilization due to inefficient measurements [9]. The second category employs explicit notifications from the network, such as Explicit Congestion Notification (ECN) [10] and Low Latency, Low Loss, and Scalable Throughput (L4S) [11], to trigger congestion control responses. While these protocols are efficient in controlling congestion, most are based on standard ECN, which can only convey Boolean information—congestion or no congestion. In other words, ECN does not notify the end hosts about the severity of the congestion. Moreover, the two ECN-related bits in the IP header are underutilized, as two bits can represent four distinct states, but in ECN, they represent only three states. A few mechanisms proposed recently, such as RCP [12] and XCP [13], signal more information about the congestion state and perform near optimally but require multiple bits per packet, making deployment challenging due to modifications needed in the standard Internet architecture, especially the IP header.

---

*Corresponding author
✉ inayat@etri.re.kr (I.A.)
[1]

Hence, is it possible to convey more information on the congestion state using the same two ECN bits without altering the TCP/IP header? Affirmatively, the Enhanced ECN (EECN) design introduced in this study demonstrates such capability. By coding two distinct congestion intensities within the IP header's ECN bits and reflecting this information through the same bits in the TCP header to the sender, EECN offers a novel approach to ECN code point interpretation. This method diverges from existing proposals that necessitate significant overhauls to both the Internet's architecture and the TCP/IP header, as EECN introduces no such changes. Instead, it calls for minor adjustments at the transport layer of the end hosts to reinterpret EECN-specific code points.

In a nutshell, the contributions of this work include:

- Introduction of EECN, a mechanism that encodes two distinct congestion levels within the network by utilizing the two ECN bits in the IP header.

- Use of the same two bits in the TCP header, originally meant for ECN support, to communicate these two congestion levels back to the sender.

- A novel approach where EECN does not require all network nodes to be EECN capable, diverging from many existing mechanisms by using the same bits to negotiate EECN capability during the handshake process.

- Application of ECN bits in control packets during the handshake to ascertain the congestion state at the connection's outset and accordingly set the initial cWnd size—this is particularly beneficial for managing short-lived network flows.

- Development of simple yet effective congestion control mechanisms based on multilevel congestion feedback from the network, demonstrating the practicality of EECN.

- Introduction of a method for intermediate routers to set congestion level bits when queue occupancy hits a specified threshold, enhancing network responsiveness to congestion.

The results of EECN are promising, reducing the packet drop rate by 70% compared to ECN [10] and 40% compared to VCP [14] and the number of packets marked by 96% compared to ECN and 76% compared to VCP. Furthermore, it significantly reduces flow completion times by 61% compared to ECN, enhancing throughput for short-lived network flows, which are predominant in IoT networks. The mechanism also lowers the RTT and end-to-end delay across all flows, as detailed in the results section.

The rest of the paper is organized as follows: We present related work in Section 2. In Section 3, an overview of standard ECN is given. We explain in detail the proposed EECN mechanism in Section 4 and its deployment and coexistence with ECN in Section 5. The results and simulation are discussed in Section 6. Section 7 concludes this paper.

## 2. Related Work

Congestion control in TCP has been under research for quite a long time. Dozens of TCP variants and AQM schemes have been proposed and standardized for different network environments [15] [16]. Cross-layer solutions, including ECN [17] and L4S [11], have been proposed, with the former being an Internet standard today. However, these solutions still come with some limitations.

There are different categories of congestion control approaches, i.e., reactive approaches based on single-bit network feedback, multi-bit feedback from the network, and end-host-based approaches. The end-host-based approaches include TCP Vegas [5] and FAST [6], which make decisions based on the RTT of the packet received. Another prominent congestion control mechanism, BBR [7], makes congestion control decisions based on the estimation of time-varying available bandwidth. These reactive protocols sometimes make inefficient decisions based on erroneous measurements of available bandwidth and RTT.

The second category of congestion control algorithms is based on single-bit feedback from the network using ECN. CoDel [18] and PIE [19] signal congestion based on the sojourn time and probability to limit the queueing delay and RTT. DCTCP [20] allows the sender to adjust its cWnd based on the degree of congestion determined from the sequence information provided by single-bit feedback. However, the degree of congestion is determined from single-bit feedback and not explicitly notified by the network. These single-bit approaches work based on estimation and calculation and may give inaccurate estimations of congestion. Kick-Ass [21] proposed a signaling mechanism to embed explicit information into TCP/IP packets and communicate between routers and endpoints. However, these kinds of approaches are discouraged by standardization bodies which require huge efforts and modification to the TCP/IP stack.

Several receiver-driven approaches have been developed to manage congestion and maintain ultra-low network queuing delay [22–24]. For instance, Homa [22] introduces a receiver-driven flow control mechanism that utilizes network priority queues to ensure minimal queuing delay. On the other hand, NDP [24] employs a receiver-driven pull mechanism for traffic control. While these approaches are effective in reducing queuing delays, it is worth noting that they are primarily tailored for data center networks and experience issues related to underutilization of network links [25].

ABC [26] and VCP [14] are closer to our work in spirit as they also use the two ECN bits with different interpretations. The authors in [26] reinterpret the ECN code points in the IP header and use these two bits to signal non-ECN capable transport, "accelerate" ECT (1), "brake" ECT (0), and standard Congestion Experienced (CE) flag. However, there exist some issues in the design and deployability. Firstly, the mechanism considers only the cellular environment where the base station keeps a per-user state. Secondly, the flag ECT (1) is used to inform routers about the ECN capability of this connection. Using this flag for any other purpose raises the question of how intermediate routers would know the ECN capability of the connection, which is not addressed in the paper. Furthermore, the "accelerate" flag in the proposed scheme and the ECT (1) flag in the standard ECN cause a similar congestion response from the sender, which is to increase the cWnd. Lastly, the mechanism uses an extra NS bit of the TCP header to echo accelerate and brake flags to the sender; however, the NS bit is never deployed and was proposed for a different purpose.

The authors in [14] also use the two ECN bits to encode four pieces of information, i.e., non-ECN capable transport, Low-load, High-load, and Overload. Nevertheless, the same issue of communicating the ECN capability to intermediate routers is overlooked. It lacks the mechanism to echo this information to the sender in the TCP acknowledgment (ACK) packet. VCP, while effective in certain scenarios, tends to converge slowly, which is not advantageous for short-lived flows.

There has also been a proposal [27], to implement ECN on SYN-ACK packets, a concept referred to as ECN+. However, the utility of ECN marking on SYN-ACK packets primarily benefits situations where the receiver rather than the sender initiates the connection. In contrast, this approach offers no advantages for short-lived flows in other typical TCP scenarios, i.e., scenario 2 and scenario 3, as depicted in Fig. 3, and Fig. 4. These scenarios illustrate that the benefits of ECN+ are not universally applicable across different TCP connection types. Another RFC [28] suggests extending ECN capabilities to control packets, an enhancement referred to as ECN++. While these proposals aim to extend ECN functionalities for control packets to reduce their drop rates and prevent prolonged delays awaiting transmission timeouts before retransmission — issues that significantly impact short-lived connections — our proposed solution goes a step further. It not only extends ECN capabilities to the SYN-ACK packet but also leverages this enhancement to improve the performance of short flows by adjusting the initial cWnd based on the current congestion conditions notified through the SYN-ACK packet, thereby offering a direct performance benefit for short-lived connections.

In contrast to the congestion control mechanisms mentioned earlier, our proposed approach makes efficient use of the existing two ECN bits in the IP header to convey additional congestion information. Additionally, it introduces a mechanism for marking TCP control packets, leading to enhanced throughput, reduced end-to-end delay by 20%, and shorter flow completion time for short-lived flows by an average of 61% compared to standard ECN and TCP New Reno, as evident from the results.

## 3. Overview of ECN

ECN is an extension to IP and TCP that allows end-to-end notification of network congestion [10]. In both the IP and TCP headers, two flags facilitate ECN support. The IP flags are ECN-Capable Transport (ECT) and Congestion Experienced (CE). The two reserved bits for ECN carry three pieces of information in the IP header. The code point (00) indicates a Not-ECT, code points (01) and (10) both signify ECT, and (11) signifies CE.

The standard ECN negotiation process is illustrated in Fig. 1. In an ECN-capable router, when a packet with an ECT (1) or ECT (0) flag is received during network congestion, instead of dropping the packet, it is marked by setting the CE flag, indicating congestion experienced. Once the receiver receives this packet, it echoes the congestion feedback back to the sender using two bits in the TCP header to inform the sender about the congestion.

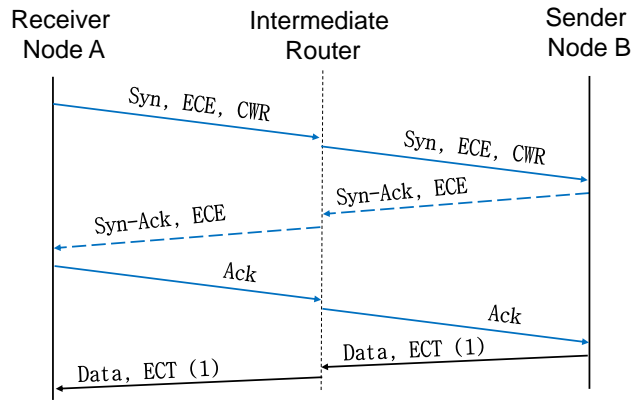

**Figure 1:** ECN negotiation and working.

**Table 1**
IP header flags for EECN support.

| IP Flags | | Explanation |
| --- | --- | --- |
| Bit 15 (CE) | Bit 14 (ECT) | |
| 0 | 0 | Not EECN Capable Transport- ECT (0) |
| 0 | 1 | EECN Capable Transport- ECT (1) |
| 1 | 0 | Congestion level 1- CL (1) |
| 1 | 1 | Congestion level 2- CL (2) |

The two flags in the TCP header for ECN support are ECN-Echo (ECE) and Congestion Window Reduced (CWR). The receiver sets the ECE bit in the corresponding ACK packet to inform the sender about congestion. The sender responds by dropping its cWnd and setting the CWR flag in the next data packet, signaling the successful reception of the ECE flag. The receiver stops setting the ECE flag after receiving the CWR flag.

During the handshake, the ECE and CWR flags are employed to negotiate ECN capability. The sender sets the ECE and CWR flags in the SYN packet to inform the receiver that the sender is ECN capable. If the receiver is also ECN capable it sets the ECE flag in the SYN-ACK packet.

## 4. Proposed EECN Mechanism

The ECN design has certain limitations, including the inability to deliver feedback about the severity of congestion. The flags ECT (0) and ECT (1) both convey the same information, which is ECN-capable transport—a situation we consider an underutilization of the two bits reserved for ECN in the IP header. Additionally, in standard ECN, SYN, and SYN-ACK packets during the handshake are not ECN-capable, although proposals exist to extend ECN capabilities to control packets, as previously discussed. The rationale behind these proposals is to reduce the drop rate of control packets only. The proposed EECN uses the same two ECN bits but notifies two distinct congestion levels. These two levels then trigger different congestion control policies by the TCP sender as explained later to avoid congestion without compromising throughput and reducing the packet loss considerably. EECN reformulates the ECN flags and the way network nodes respond to them. The details of EECN flags are explained in the next section.

### 4.1. EECN Flags

When a packet reaches a router, it marks or drops it based on the implemented queue management algorithms. Marking involves setting the ECN bits in the IP header of that packet. The IP header designates bit 14 as ECN Capable Transport (ECT) and bit 15 as Congestion Experienced (CE) for ECN. There are four code points: (00, 01, 10, and 11 ). In EECN, these code points are interpreted differently, as illustrated in Table 1.

**Table 2**
TCP header flags for EECN support.

| TCP Flags | | | Explanation |
|---|---|---|---|
| Bit 9 (ECE) | Bit 8 (CWR) | Syn/Ack | |
| 0 | 0 | 1/0 or 1/1 | ECT (0) Not EECN Capable Transport |
| 0 | 1 | 1/0 | ECT (1) – EECN Capable Transport |
| 0 | 1 | 1/1 or 0/1 | ECT (1) – EECN Capable Transport |
| 1 | 0 | 1/1 or 0/1 | CL (1) - Congestion level 1 |
| 1 | 1 | 1/1 or 0/1 | CL (2) - Congestion level 2 |
| 0 | 1 | 0/0 | CWR- Congestion Window Reduced |

In the standard ECN mechanism, ECT (0) and ECT (1) convey the same information, leading to the underutilization of two essential bits in the IP header. Therefore, EECN redefined and reinterpreted these flags. The new interpretation allows for the indication of two levels of congestion. Specifically, code point (00) indicates that the connection carries packets from a non-ECN-capable transport (ECT (0)), while (01) represents packets from an ECN-capable transport (ECT (1)). Code points (10) indicate the detection of Congestion Level 1 (CL (1)), and (11) denotes Congestion Level 2 (CL (2)) as illustrated in Table 1. More details about these congestion levels can be found in Section 4.3.

In standard ECN, when a packet marked with CE is received, the receiver notifies the sender about the congestion event by setting an ECE flag in the TCP header of the corresponding ACK packet. EECN modifies these flags, allowing the receiver to communicate both CL (1) and CL (2) congestion information back to the sender. It also enables marking in control packets during the handshake phase using the same flags but with different code points.

Similarly in ECN, the two ECE and CWR flags in the TCP header convey information about ECN-capable transport and non-ECN-capable transport, ECN-echo, and congestion window reduction. EECN redefines the interpretation of these code points, particularly in conjunction with SYN and ACK flags. This reinterpretation serves two purposes. Firstly, it allows for marking on SYN and SYN-ACK packets during the handshake. Secondly, it enables an endpoint to communicate more detailed information to the other end. For example, in traditional ECN, the code point (01) indicates congestion window reduction. In EECN, however, (01) signifies ECN-capable transport when set in SYN, SYN-ACK, or ACK packets, but denotes congestion window reduction when set in data packets (i.e., when both SYN and ACK bits are not set). The code points (10) and (11) indicate CL (1) and CL (2), respectively, and are used in ACK and SYN-ACK packets. The EECN TCP flags are described in Table 2.

### 4.2. EECN Mechanism

During the initial handshake, the sender and receiver both employ the code points (00) to signify a non-ECT and (01) for an ECT in the SYN and SYN-ACK packets. Unlike standard ECN, EECN also sets the code point (01) (ECT (1)), in the IP header of SYN and SYN-ACK packets. This allows intermediate routers to mark existing congestion. When a SYN or SYN-ACK packet with a congestion flag—either CL (1) or CL (2)—is received, the host acknowledges the congestion by echoing it back to the sender in the corresponding SYN-ACK or ACK packet. The sender then adjusts its initial cWnd based on the detected level of congestion. If the marked packet was a data packet during the data exchange phase, the sender reduces its cWnd in response to the congestion notification. The sender then includes the CWR flag in the TCP header of the subsequent data packet to inform the receiver about the measures taken in response to the notified congestion. Fig. 2, Fig. 3, and Fig. 4 illustrate the TCP handshake and the exchange of data and ACK packets within EECN.

#### *4.2.1. EECN Enabled Control Packets*

The standard ECN lacks the capability to mark control packets. However, we contend that enabling this feature to control packet marking offers significant advantages. It allows the TCP sender to make more informed congestion control decisions right at the start of a connection. For example, short-lived network flows often last only a few seconds can benefit from this. Consider a 16KB flow: it could be transmitted in approximately 16 segments if the size of each TCP segment is 1000 Bytes. These short flows typically conclude before the TCP reaches its optimal congestion window size.

EECN addresses this by enabling SYN and SYN-ACK packets to carry the ECT (1) flag in the IP header during the handshake phase. If these packets encounter congestion along the network path, they are marked with a congestion

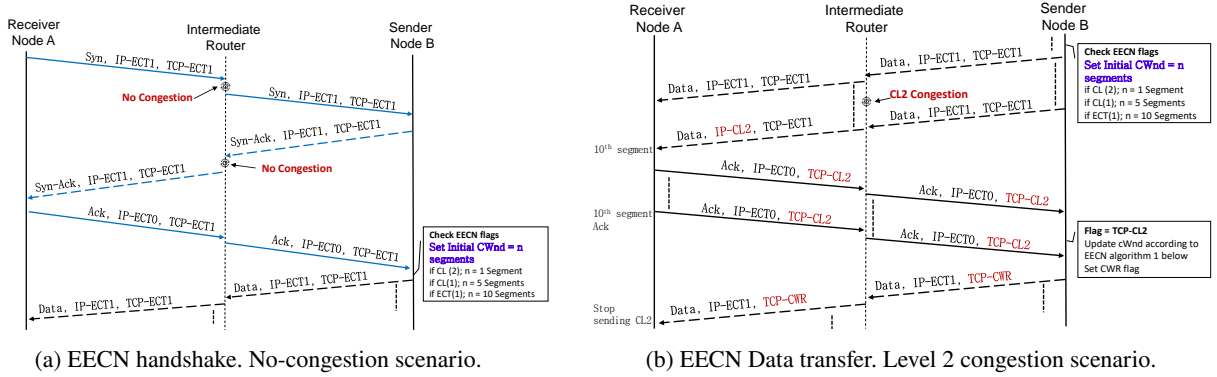

(a) EECN handshake. No-congestion scenario.  (b) EECN Data transfer. Level 2 congestion scenario.

**Figure 2:** EECN mechanism in a scenario where the receiver node requests the connection.

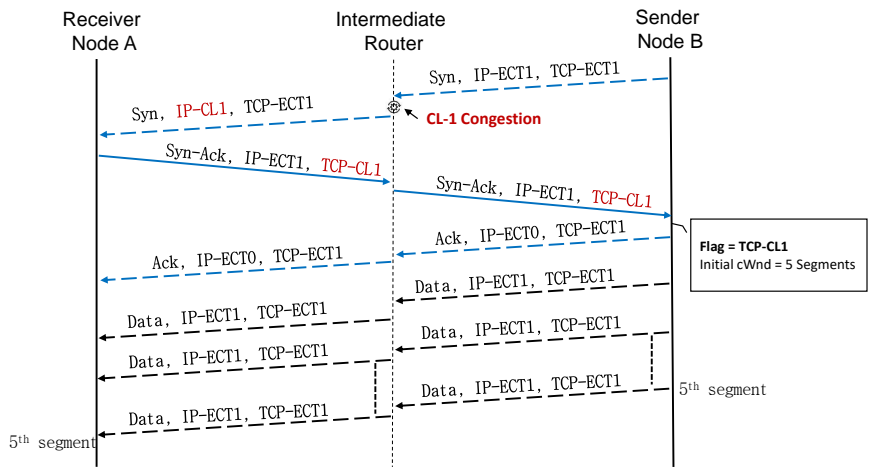

**Figure 3:** EECN mechanism in a scenario where the sender node requests the connection.

flag, just like other data packets in EECN-capable transport. Should a SYN packet be marked with either the CL (1) or CL (2) flag, this congestion information is echoed back to the sender in the TCP header of the SYN-ACK packet. Similarly, if the SYN-ACK packet is marked with either CL (1) or CL (2), the congestion event is relayed back to the sender of the SYN-ACK in the TCP header of the corresponding ACK packet. This mechanism is detailed in Fig. 2, Fig. 3, and Fig. 4.

EECN sets the initial cWnd based on the network conditions observed during the handshake phase. For instance, the initial cWnd is set to ten segments if no congestion is detected, five segments if congestion level 1 is encountered, and one segment if the control packets experience congestion level 2. This adaptation leads to notable performance improvements for short-lived network flows, as demonstrated by the results. The EECN mechanism is further elucidated through three possible TCP scenarios below.

*Scenario 1:* The connection establishment process begins when the receiver node A sends a SYN packet to the sender node B, as depicted in Fig. 2. In this SYN packet, node A sets both IP-ECT (1) and TCP-ECT (1) flags. There is no congestion in the network path (i.e., intermediate router), therefore the packet is received without marking. The IP-ECT(1) flag indicates to node B that the packet has not encountered congestion along its path. Meanwhile, the TCP-ECT (1) flag informs node B that the connection is EECN capable. If node B is not EECN capable, it would respond by setting both TCP-ECT (0) and IP-ECT (0) in the SYN-ACK packet, signaling to the sender that the receiver does not support EECN. In such a case, the sender would refrain from setting IP-ECT (1) on subsequent data packets. However, in the examples discussed here, all nodes are EECN capable, so node B sets both IP-ECT (1) and TCP-ECT (1) in its SYN-ACK packet. As illustrated, the SYN-ACK packet also encounters no congestion and is forwarded without marking.

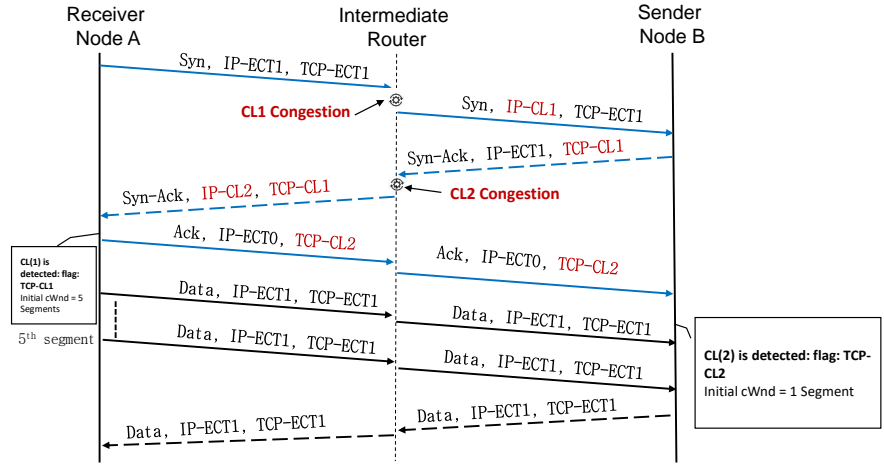

**Figure 4**: EECN mechanism in a scenario where the receiver node requests the connection in a bidirectional communication.

Following this, node A sends an ACK packet back to node B. Upon receiving this ACK packet with the TCP-ECT (1) flag, node B sets its initial cWnd to ten segments and sends out ten segments. The eighth packet among these encounters level 2 congestion at an intermediate router, which marks the packet by setting the IP-CL (2) flag. Node A acknowledges this congestion event by echoing it in the TCP header of the subsequent ACK packet. After receiving this ACK packet, node B adjusts its cWnd as per Algorithm 1 and sets the CWR flag in the TCP header of the next data packet. Node A ceases to set the TCP-CL (2) flag once it receives the CWR flag.

*Scenario 2:* In Scenario 2, sender node B initiates the connection by sending a SYN packet and sets both IP-ECT (1) and TCP-ECT (1) flags. This is illustrated in Fig. 3. The SYN packet encounters congestion level 1 at an intermediate router. As a result, the router marks the packet with the IP-CL (1) flag, indicating the presence of congestion.

The sender node B is then informed of this congestion incident through the SYN-ACK packet received from the receiver node. Upon recognizing the congestion level 1 experienced by the SYN packet it has sent, node B adjusts its initial cWnd to five segments. This is a response to the network conditions it has observed, aligning with the protocol's aim to manage congestion effectively. Node B sends an ACK packet to confirm the successful reception of the SYN-ACK packet. Subsequently, it transmits five data segments, adhering to the initial cWnd set at five, which reflects the network's congestion level as encountered by the SYN packet.

*Scenario 3:* In Scenario 3, shown in Fig. 4, both end nodes, receiver node A and sender node B, act simultaneously as client and server, exchanging data. Node A initiates the connection, but in this scenario, both the SYN and SYN-ACK packets encounter congestion during the handshake phase.

Node B sends a SYN packet, which experiences congestion level 1 at an intermediate router. This leads to the router setting the IP-CL (1) flag. Node B then communicates this congestion event to node A in the SYN-ACK packet by setting the TCP-CL (1) flag. Additionally, it sets the IP-ECT (1) flag in the SYN-ACK packet. However, the SYN-ACK packet itself faces congestion level 2 at the router, resulting in the router marking it with the IP-CL (2) flag. Consequently, the SYN-ACK packet received by node A carries both the IP-CL (2) flag, indicating the congestion it experienced, and the TCP-CL (1) flag.

The TCP-CL (1) flag here conveys two pieces of information: *firstly, that the receiver node B is EECN capable, and secondly, that the SYN packet it sent earlier encountered congestion level 1.* As a result, node A sets its initial cWnd to five segments. Node A then sends an ACK packet, signaling the reception of the SYN-ACK packet and the completion of the handshake. In this ACK packet, node A includes the TCP-CL (2) flag to echo back the congestion level 2 experienced by node B's SYN-ACK packet. Upon receiving this ACK packet with the TCP-CL (2) flag, node B sets its initial cWnd to one segment.

**Table 3**
EECN thresholds selection for congestion marking.

| Threshold | Throughput (Mbps) | | Packet Drop | E-to-E Delay (ms) | |
|---|---|---|---|---|---|
| | EF | SF | Total | EF | SF |
| $Th_1 = 0.5, Th_2 = 0.7$ | 48.7 | 16.1 | 425 | 2.5 | 3.0 |
| $Th_1 = 0.3, Th_2 = 0.5$ | 48.6 | **16.1** | 462 | **1.9** | **2.7** |
| $Th_1 = 0.2, Th_2 = 0.4$ | 46.8 | 12.7 | 614 | 1.86 | 2.5 |

### 4.3. EECN Packet Marking

When a router receives a packet marked with an ECT (1) flag in the IP header, it assesses the level of congestion at that router based on the state of the network over the subsequent 100 milliseconds. We formulate two distinct levels of congestion: congestion level 1 (CL(1)) and congestion level 2 (CL(2)).

The determination of the congestion level is based on the queue occupancy, which is calculated using equation 1. If the calculated queue occupancy is more than 30% but less than 50% within the next 100 ms, as calculated using equation 1, the congestion is classified as level 1 (CL(1)). On the other hand, if the queue occupancy exceeds 50%, the congestion is deemed to be at level 2 (CL(2)). Since $\gamma$ and $\alpha$ are the arrival and departure rates of packets per second, respectively, the term $(\frac{\gamma}{10} - \frac{\alpha}{10})$ in Equation 1 is used to estimate the change in queue size based on the packet arrival rate ($\gamma$) and departure rate ($\alpha$) over a 100 ms time window. This formulation captures the short-term dynamic effects of traffic, providing insight into how the queue grows or shrinks based on traffic conditions. Specifically, if the arrival rate exceeds the departure rate, the queue begins to fill up, signaling a higher level of congestion. Conversely, if the departure rate is higher than the arrival rate, the queue shrinks, indicating reduced congestion. This mechanism helps predict the future state of the queue and thus the congestion level, reflecting the temporal and dynamic nature of traffic flows. While this model provides an efficient and straightforward method for estimating congestion, it does not account for certain traffic behaviors such as burstiness or variations in packet inter-arrival times. Future work could expand on this approach by incorporating more detailed models that better reflect the complexities of real-world traffic patterns.

$$\text{Congestion Level} = CL = \frac{\text{Current queue size} + (\frac{\gamma}{10} - \frac{\alpha}{10})}{\text{Total length of the queue}} \quad (1)$$

Where

$\gamma$ = Arrival rate of packets per second
$\alpha$ = Departure rate per second
$Th_1$ = Threshold 1 = *0.3*
$Th_2$ = Threshold 2 = *0.5*

Case 1: if $(Th_2 > CL < Th_1)$
- No congestion

Case 2: if $(Th_2 > CL \geq Th_1)$
- Congestion level 1, EECN flag is $CL(1)$

Case 3: if $(Th_2 \leq CL)$
- Congestion level 2, EECN flag is $CL(2)$

For example, if the current queue size at a router is 60 packets, and the total queue capacity is 100 packets, and imagine the packet arrival and departure rate is the same, then according to the earlier mentioned equation, the calculated congestion level (CL) in this case is 0.6. This value falls within the range specified for congestion level 2. Therefore, the router identifies this as a congestion level 2 situation and sets the CL (2) flag, represented by the code point (11), on the packet before forwarding it. The threshold values of 0.3 and 0.5 are chosen for their strategic balance. A lower threshold might lead to excessive packet marking and packet dropping when there is still ample space in the queue, as demonstrated by the simulation results in Table 3. Here, "EF" represents elephant flows, which are long FTP-like flows, and "SF" stands for short-lived flows. On the other hand, a higher threshold could lead to longer queuing delays and an increased risk of queue overflow, particularly in cases of sudden data flow surges at the router. Thus, these mid-range threshold values are selected to avoid both excessive packet marking and packet drop that may be caused by potential queue overflow.

When a packet arrives at a router, the first step is to check its flags. If the CL (2) flag is already set, indicating that another router in the network path has identified high-level congestion, the packet is simply forwarded without any changes. If the packet carries the CL (1) flag, the router evaluates its own queue occupancy based on equation 1. If the congestion is below a certain threshold $th_2$, the packet is forwarded as is, retaining the existing CL (1) flag. However, if the router's congestion meets or exceeds $th_2$, it updates the flag to CL (2) before forwarding the packet.

In cases where a packet arrives with the ECT (1) flags set and no congestion level flags, the router assesses its queue occupancy. Based on this evaluation, the router decides whether to set a CL (1) or CL (2) flag or forward the packet without setting any additional congestion flags.

### 4.4. Congestion Control in EECN

The TCP sender implements various congestion control mechanisms in response to the congestion flags received in the TCP header. Our proposed EECN introduces a more efficient congestion control mechanism. EECN adapts to network conditions more effectively by responding to the specific congestion level flags, CL (1) and CL (2), communicated by intermediate routers. This targeted response allows for a finer adjustment of the sender's cWnd and optimizing data flow based on current network congestion levels. By dynamically adjusting its transmission rate following the feedback received via these flags, EECN enhances overall network efficiency,

#### 4.4.1. Initial cWnd

In the EECN, the initial setting of the cWnd by the TCP data sender is directly influenced by the flags received during the handshake phase. Specifically, when the first ACK packet is received with the ECT (1) flag, the sender assumes there is no congestion in the network path and accordingly sets the initial cWnd to ten segments. However, if the received flag is CL (1), indicative of a mild level of congestion, the initial cWnd is conservatively set to just five segments. This approach is particularly effective for handling short-lived network flows. On the other hand, the presence of a CL (2) flag, signaling more severe congestion, prompts a stringent flow control response, wherein the initial cWnd is limited to one segment. This tiered response strategy ensures efficient and adaptive management of network traffic under dynamic network conditions.

#### 4.4.2. Updating cWnd

We have developed a straightforward yet distinct method for adjusting the cWnd based on the level of congestion indicated during the data exchange phase. When the CL (2) flag is received, signaling severe congestion, the cWnd reduction strategy is twofold. If the current RTT exceeds the average RTT for the connection, the cWnd is reduced by a factor of $d = 8$. If the current RTT is less than or equal to the average, the reduction is more moderate, at a factor of $d/2$. After this reduction, if the resultant cWnd is smaller than two segments, it is automatically adjusted upward to two segments. The divisor $d$ used for reducing the cWnd is chosen for its strategic balance. Moreover, the primary objective of this research is to emphasize the effectiveness of multilevel congestion feedback. Implementing a straightforward approach by scaling down the cWnd with different factors for congestion level 1 and congestion level 2 serves to validate the proposed two-bit multilevel congestion feedback's efficacy. Employing a more complex response might obscure the clear demonstration of this mechanism's efficiency.

In scenarios where the CL (1) flag is received, indicating milder congestion, the cWnd adjustment is governed by the decay function shown in equation 2. This function is applied only if the current RTT for the connection is greater than the average RTT. If the current RTT does not exceed the average, the cWnd remains unchanged. In this case, the receiver continues to send the TCP-CL (1) flag in the subsequent three ACKs before stopping, unlike other congestion conditions where the receiver ceases sending TCP-CL flags after receiving the TCP-CWR flag from the sender.

$$cWnd = cWnd \cdot e^{-\beta} \tag{2}$$

**Algorithm 1** Processing EECN Congestion Notification by Sender Node
---
**Input:** Packet with EECN bits
    CL (1): TCP header congestion level 1
    CL (2): TCP header congestion level 2
    ECT (1): EECN capable transport
**Output:** cWnd
1: **if** CL (2) **then**
2:   **if** cRTT ≥ avgRTT **then**
3:     $SsThreshold \leftarrow max(2, cWnd/d)$
4:     $cWnd \leftarrow SsThreshold$
5:   **else**
6:     $SsThreshold \leftarrow max(2, cWnd/(d/2))$
7:     $cWnd \leftarrow SsThreshold$
8:   **end if**
9: **end if**
10: **if** CL (1) **then**
11:   **if** cRTT ≥ avgRTT **then**
12:     $\beta \leftarrow cRTT - avgRTT$
13:     $SsThreshold \leftarrow cWnd \times e^{-\beta}$
14:     $cWnd \leftarrow SsThreshold$
15:   **else**
16:     $cWnd \leftarrow cWnd$ (*no change*)
17:   **end if**
18: **end if**
19: **if** ECT (1) **then**
20:   $cWnd \leftarrow cWnd$ (*no change*)
21: **end if**
22: **return** $cWnd$

The $\beta$ is the factor by which the cWnd is decayed and it is defined as the difference between the current RTT (cRTT) and the average RTT (avgRTT), i.e., $\beta = cRTT - avgRTT$. The factor $\beta$ represents the difference between the current RTT and the average RTT observed during the connection. This difference reflects the extent of queuing delay in the network, which directly correlates with congestion. As queuing delay increases, the current RTT grows larger than the average RTT, signaling congestion. By using $\beta$ as the decay factor, we ensure that the cWnd adjustment is proportional to the degree of congestion, allowing the sender to react more conservatively in highly congested environments.

The use of exponential decay in this context provides a smooth and responsive adjustment of the congestion window. By scaling the cWnd based on the factor $e^{-\beta}$, the sender can reduce its transmission rate in a manner proportional to the observed congestion, thereby avoiding sudden drops in throughput. This approach allows for more gradual and controlled reductions in cWnd as the network experiences congestion, ensuring that flows remain stable while adapting to changing conditions.

### 4.4.3. Slow-Start phase

We have also refined the slow start mechanism in TCP by moderating the increase of the cWnd during this phase, taking into account the RTT. The algorithm enters the slow-start phase when the current cWnd is below the slow-start threshold; otherwise, it shifts to the congestion avoidance phase. In the slow-start phase, if $\beta$ (the difference between current and average RTT) is positive, the cWnd increment per ACK received is scaled down. Specifically, it is increased by one segment multiplied by "$\sigma$" ($\sigma = 0.3$) per ACK received where "$\sigma$" is the smoothing factor ($0 < \sigma \leq 1$). if $\beta$ is not positive, the conventional TCP New Reno slow-start approach is employed. This adjustment effectively moderates the rate of increase in cWnd in scenarios of severe congestion, as indicated by the RTT.

**Algorithm 2** Slow Start (SS) and Congestion Avoidance (CA) Algorithm in EECN

**Input:** Packet with EECN bits
    CL (1): TCP header congestion level 1
    CL (2): TCP header congestion level 2
    ECT (1): EECN capable transport
    SZ: Segment Size
**Output:** cWnd
1: **if** cWnd < SsThreshold **then**
2:    *Slow Start*
3:    **if** cRTT $\geq$ avgRTT **then**
4:       $cWnd\mathrel{+}= SZ \times \sigma$ *where* $1 \geq \sigma > 0$
5:    **else**
6:       $cWnd\mathrel{+}= SZ$
7:    **end if**
8: **else**
9:    *Congestion Avoidance*
10:    **if** cRTT $\geq$ avgRTT **then**
11:       $\beta \leftarrow cRTT - avgRTT$
12:       $cWnd \leftarrow cWnd \times e^{-\beta}$
13:    **else**
14:       $cWnd\mathrel{+}= SZ \times \sigma$ *where* $1 \geq \sigma > 0$
15:    **end if**
16: **end if**
17: **return** $cWnd$

### *4.4.4. Congestion Avoidance*

In our redesigned congestion avoidance strategy, we also incorporate RTT considerations. Instead of simply incrementing the cWnd modestly as in TCP New Reno, we apply the aforementioned decay function when $\beta$ is positive. If $\beta$ is negative, we increment the cWnd slightly, employing a smoothing factor "$\sigma$" set to 0.02 in the congestion avoidance phase.

Algorithm 1 explains how the sender node in the EECN handles packets marked with EECN flags after congestion is detected. At the start of a connection, the sender node enters the slow-start phase, increasing the cWnd as detailed in Algorithm 2 until the cWnd exceeds the slow-start threshold. Once this threshold is exceeded, the connection transitions into the congestion avoidance phase. Algorithm 2 outlines the steps for both the slow-start and congestion avoidance phases in EECN. EECN flags indicating congestion can be received at any time during the connection. When these flags are received, the connection enters the congestion detection phase and updates its cWnd as described in Algorithm 1.

Given EECN's capability to signal two distinct levels of network congestion, we argue that it lays the groundwork for crafting more effective congestion and flow control mechanisms. These mechanisms are particularly crucial in meeting the demanding bandwidth and low latency requirements of modern network applications. This straightforward congestion and flow control mechanism underscores the efficiency of EECN's multilevel network feedback system.

## 5. Deployability and Compatibility

We address the deployability of EECN. EECN requires no modifications to the TCP/IP headers and allows it to coexist seamlessly with traditional ECN. This compatibility facilitates incremental deployment. We thoroughly examine potential challenging scenarios to affirm the successful coexistence of EECN with ECN.

### 5.1. EECN Router

The approach to packet marking at intermediate routers in EECN is similar to traditional ECN, but EECN requires routers to mark two levels of congestion based on queue occupancy. This requirement poses no significant challenge in the context of today's software-driven network environments, and no additional changes at the network layer are necessary. If an ECN router receives a packet with a non-ECT IP header flag (code point (00)), both ECN and EECN routers treat this as a connection from a non-ECN-capable transport. Conversely, if a packet with an ECT (1) (code point (01)) or ECT (0)/CL (1) (code point (10)) IP flag is received, it is considered a markable packet by both ECN and EECN routers as explained in Fig. 5. In situations of congestion, an ECN router will change the IP header flag to CE, i.e., code point (11), which corresponds to the CL (2) flag in EECN. Upon receiving this flag, an EECN sender activates its congestion control response to the CL (2) flag while an ECN sender activates its congestion response to the CE flag. When a code point (10) is received; an ECN router interprets this as a markable packet, whereas EECN views it as a markable packet from an ECN-capable transport that has encountered congestion level 1. Nonetheless, both ECN and EECN routers process this packet similarly by checking the congestion level and marking or ignoring the packet as required. However, this issue is non-existent in EECN-enabled connections, as an EECN-enabled sender sets the code point (01) (ECT (1) flag) only instead of code point (10) (ECT (0) in ECN) to indicate EECN-capable transport. As illustrated in Fig. 5, an EECN router marks packets with either CL (2) or CL (1). CL (2) is equivalent to CE in ECN, and CL (1) corresponds to ECT (0) in ECN. This marking scheme ensures that ECN receivers can correctly interpret these flags without any issues, even when the packets pass through an EECN router en route. Similarly, in the case of an EECN connection, an ECN router will mark packets with high congestion (code points (11)) if such congestion is detected en route. This marking ensures that EECN end hosts can correctly interpret these signals of high congestion indicated by CL (2) (code point (11)). This compatibility feature allows EECN hosts to effectively respond to congestion scenarios, even when the congestion is marked by an ECN router.

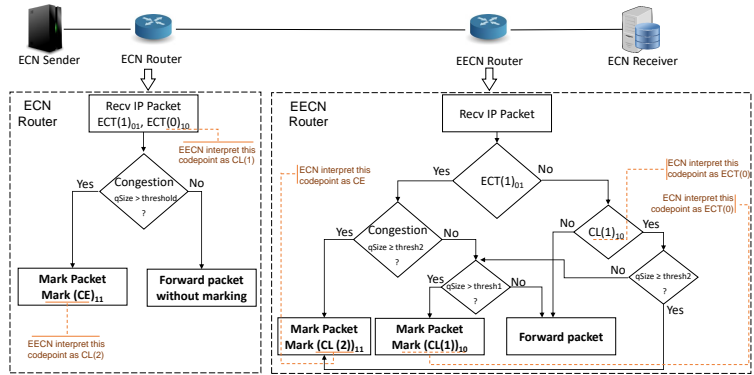

**Figure 5**: EECN coexistence with ECN.

## 6. Evaluation and Discussion
### 6.1. Simulation
We implemented the EECN mechanism in the NS-3 simulator [29], a widely used open-source simulator for network research. The simulation scenarios comprise a dumbbell network topology as shown in Fig. 6a and a multi-hop network scenario with both ECN and EECN router en route as shown in Fig. 6b, consisting of three clients, three servers, and intermediate routers with RED queue discipline [30] in packet mode having a queue size of 100 packets. The two intermediate routers are interconnected through a 100 Mbps bottleneck link. For the RED queue discipline, the minimum queue size threshold is 30, and the maximum threshold is 60. There are a total of eight concurrent flows with the same distribution as the realistic one, including two FTP elephant flows of size 1Gb, and the remaining six flows are web searches with sizes of 9KB and 16KB, representing short-lived network flows as over 50% of web searches are of size less than 10 KB [25]. In the results, "EF" denotes Elephant Flows, signifying the long FTP flows, while "SF" refers to Short Flows, representing short web requests. We compared the results with TCP New Reno [27] with ECN disabled, TCP New Reno with ECN enabled, and VCP [14]. The VCP parameters are selected from its original paper [14]. We have chosen New Reno as it is a widely used TCP variant. For the EECN, TCP New Reno is employed with optimizations in the Slow-start and congestion avoidance algorithm and maintaining the cWnd based on the multilevel feedback from the network, as explained in the previous section.

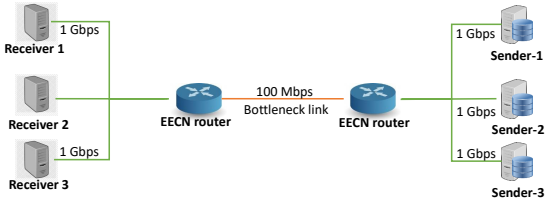
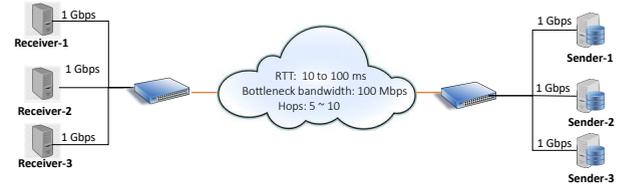

(a) Dumbbell network topology with EECN routers.

(b) Multi-hop network scenario with ECN and EECN routers.

**Figure 6:** Network scenarios with a bottleneck link.

**Table 4**
Comparison of dropped and marked packets.

|  | Packets Dropped | Bytes Dropped | Packets Marked | Bytes Marked | Packets Drop (%) |
| --- | --- | --- | --- | --- | --- |
| TCP New Reno | 10096 | 9861664 | - | - | 0.3 |
| ECN | 1553 | 1217468 | 8275 | 8697492 | 0.055 |
| VCP | 765 | 664680 | 1326 | 1393626 | 0.026 |
| **EECN** | **462** | **405064** | **318** | **330144** | **0.016** |

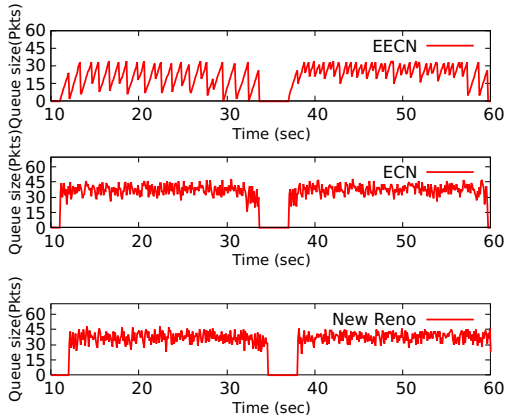
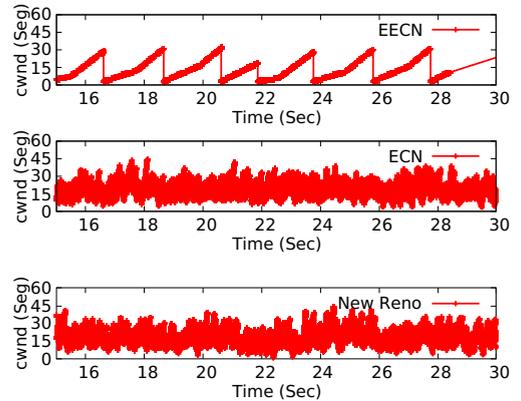

(a) Bottleneck queue size over time.

(b) cWnd of long FTP flow over time.

**Figure 7:** Bottleneck queue size and cWnd analysis over time.

## 6.2. Results and Discussion

The key performance metrics for ECN, VCP, and EECN include the percentage reduction in packet drops and packets marked at the bottleneck queue. A bottleneck queue in a network is a packet buffer at a node where the incoming traffic surpasses the node's processing capacity, resulting in an overloaded packet queue. This bottleneck queue can significantly degrade network performance, increasing latency and potentially leading to packet loss if the queue's capacity is exceeded. In the simulation scenario, the bottleneck queue is located at the EECN routers. As demonstrated in Table 4, EECN significantly outperforms the other schemes in reducing packet drops — by as much as 70% compared to ECN, 95% compared to TCP New Reno with ECN disabled, and 40% compared to VCP. Additionally, EECN marks 96% fewer packets than ECN, and 76% fewer packets than VCP. This reduction in packet marking by EECN suggests that it effectively keeps more packets within the minimum queue size threshold. This, in turn, results in fewer packet drops and lower RTT due to reduced contention in network queues.

The major reason behind the performance improvement is the ability of EECN to identify the severity of the congestion due to EECN multilevel congestion feedback. This allows EECN to adjust the flow by precisely controlling the cWnd and maintaining minimal queue occupancy, as shown in Fig. 7a. When the queue size reaches a threshold indicating congestion level 2, and RTT exceeds the average due to increased queuing, EECN responds by significantly reducing the cWnd. This action keeps queue occupancy below the threshold, leading to 96% and 76% fewer packets being marked compared to ECN and VCP, respectively.

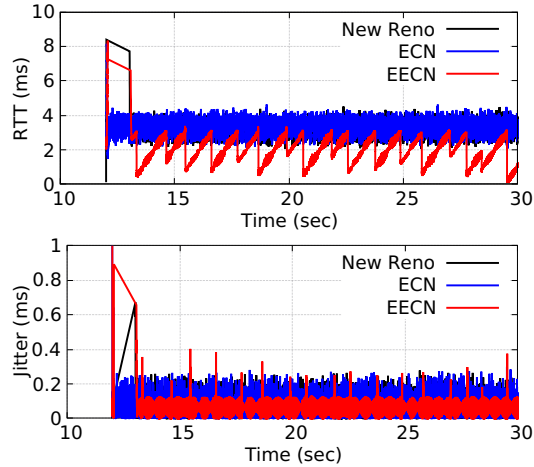

**Figure 8:** RTT and Jitter of FTP flow.

Moreover, EECN exhibits a smoother increase in queue size relative to ECN and New Reno, thanks to its smooth cWnd increment resulting from multilevel congestion feedback. This behavior is further demonstrated in Fig. 7b. In EECN, the cWnd escalates gradually during the congestion avoidance phase, leading to a smoother curve, unlike in ECN and New Reno where cWnd fluctuations are more pronounced. The sharp reduction in cWnd in response to the CL (2) flag, as observed in Fig. 7b, also impacts the queue size, causing a corresponding abrupt decrease as depicted in Fig. 7a.

The significant reduction in cWnd, triggered by congestion level 2, is clearly visible in Fig. 7b, where it creates a noticeable ripple-like pattern. This substantial decrease in cWnd, aimed at managing high congestion levels, also contributes to improved RTT performance. The top plot in Fig. 8 compares the RTT over time for EECN, ECN, and New Reno. It's observable that EECN maintains a lower average RTT compared to the other two. This is attributed to EECN's strategy of keeping the RTT below the connection's average RTT. Additionally, the gradual increase in RTT with EECN could be a result of the implemented window decay function and the slowing down of cWnd growth by the factor $\sigma$.

In contrast, ECN and New Reno display more irregular and fluctuating RTT patterns, indicative of less stable connections. The bottom plot in Fig. 8 presents a comparative analysis of jitter among these protocols. EECN exhibits lower jitter compared to its counterparts, as shown in the figure. This reduction in jitter is indicative of a more reliable and stable connection, highlighting another advantage of the EECN mechanism. The mechanism significantly improves the efficiency of short-lived network flows. Unlike ECN, EECN incorporates congestion control-related flags in both the IP and TCP headers of the SYN and SYN-ACK packets. This design allows the sending node to adjust its initial cWnd based on the network's congestion state as assessed during the handshake phase, prior to the commencement of actual data transfer.

The initial cWnd is crucial for short-lived flows, especially those lasting less than a second. Setting a high initial cWnd without prior knowledge of network conditions can sometimes exacerbate network congestion. However, EECN's approach of gauging network conditions before actual data transmission allows it to maintain a high initial cWnd in congestion-free scenarios. This results in lower Flow Completion Times (FCT) and higher throughput for such flows, as evidenced in Fig. 9 and Fig. 10.

Fig. 9 demonstrates that the FCT of elephant flows remains largely unaffected by EECN in both dumbbell and multi-hop network scenarios. However, the subplot for short-lived flows reveals a significant improvement in FCT when compared to ECN, New Reno, and VCP. The 61% reduction in flow completion time for short-lived flows can be attributed to EECN's strategy of setting a high cWnd during the TCP handshake phase, which is tailored based on the real-time network conditions communicated via EECN bits. Given that short-lived flows typically encompass only a few packets, a higher initial cWnd permits a larger volume of data to be sent immediately after the connection

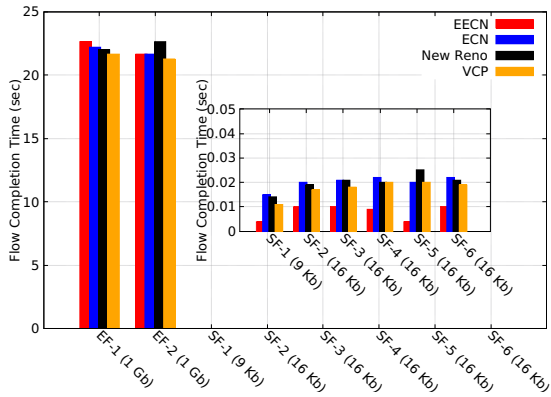
(a) FCT in dumbbell network scenario.

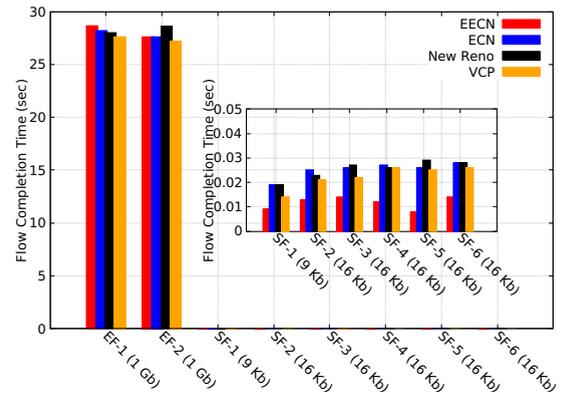
(b) FCT in multi-hop network scenario.

**Figure 9:** Flow Completion Time (FCT) of long FTP flows and short-lived web searches.

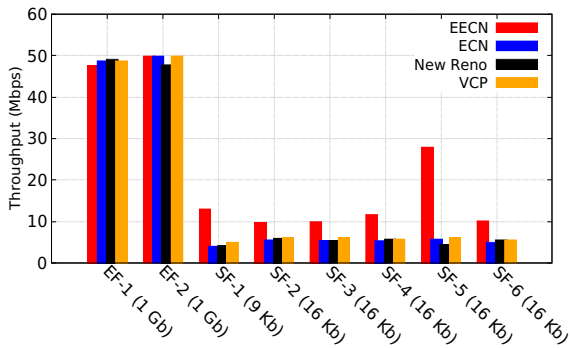
(a) Throughput in dumbbell network scenario.

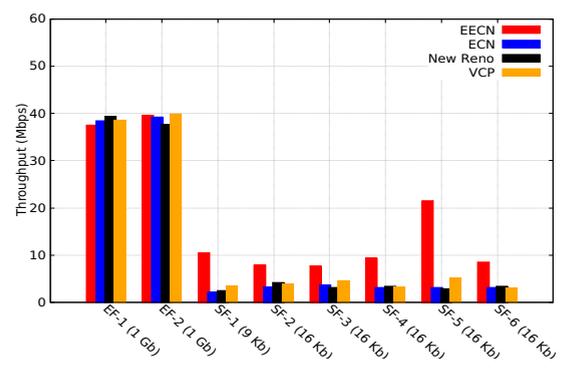
(b) Throughput in multi-hop network scenario.

**Figure 10:** Average throughput of long FTP flows and short-lived web searches.

establishment, thus markedly enhancing their throughput and reducing the time to flow completion. In contrast, the performance of long flows is not affected by this mechanism because the advantage of a higher initial cWnd diminishes over time as the flow progresses, due to the averaging effect of cWnd adjustments throughout the life of the connection. EECN dynamically adapts to changes in network traffic. In the event of a sudden traffic burst causing high congestion, EECN signals congestion level 2, prompting the sender to reduce its transmission rate. When congestion subsides, EECN no longer signals congestion, allowing the sender to gradually increase its transmission rate and ensure efficient performance under varying traffic conditions.

Similarly, Fig. 10 illustrates that short-lived flows achieve significantly higher throughput under EECN due to the same reason. While the throughput for elephant flows remains consistent, there is a substantial increase in the throughput of short-lived network flows. The throughput achieved by SF-5 is very high compared to the other short flows. On further investigating, we found that the initial cWnd of SF-5 was set to ten segments due to network conditions at the time compared to five segments for other short-flows. This adjustment reflects EECN's capability to dynamically optimize cWnd in response to real-time network conditions, significantly enhancing throughput for flows under less congested circumstances. This approach highlights EECN's effectiveness in adapting to varying network states to improve data transmission efficiency for different types of traffic.

The advantage EECN offers to short-lived flows is particularly notable when compared to their performance under ECN, TCP New Reno, and, VCP. In these traditional schemes, most short flows, typically comprising only a few packets, conclude before the end of the TCP slow start phase, failing to reach their optimal cWnd. This often results in higher FCT and lower throughput for short flows.

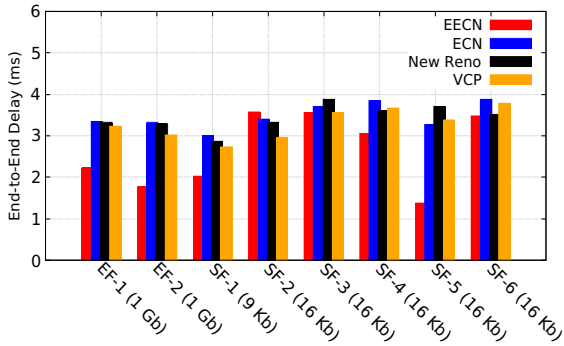
(a) End-to-End delay in dumbbell network scenario.

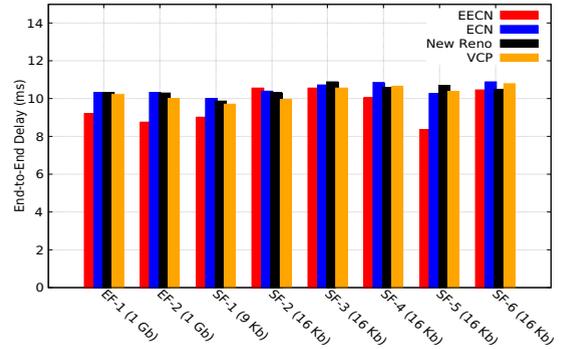
(b) End-to-End delay in multi-hop network scenario.

**Figure 11:** End-to-End delay of long FTP flows and short-lived web searches.

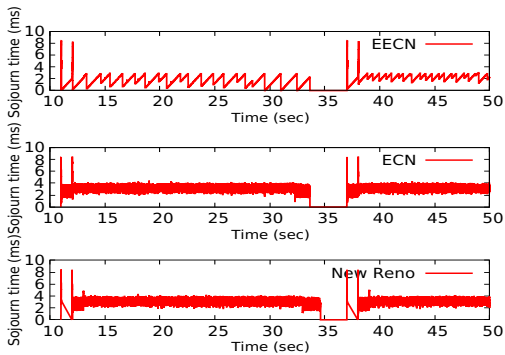
(a) Packet sojourn time in bottleneck queue.

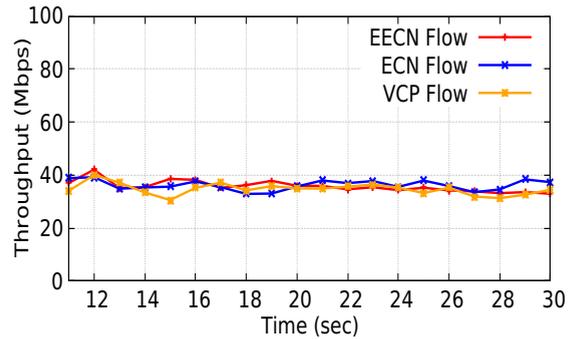
(b) Fairness among competing flows.

**Figure 12:** Packet sojourn time in bottleneck queue and fairness among competing flows.

Lastly, we evaluated the end-to-end delay across all network flows. As depicted in Fig. 11, EECN consistently achieves lower end-to-end delays compared to ECN, New Reno, and, VCP. The trend of end-to-end delay is similar in both the dumbbell and multi-hop network scenarios. Although the average delay in VCP is lower than in ECN and New Reno, the improvement is not significant. This is because VCP continues to increase the cWnd additively even after entering the high-load zone, resulting in a higher average delay compared to EECN. This EECN's performance gain is primarily due to EECN's strategy of significantly reducing the cWnd only in response to congestion level 2, followed by a gradual increase while taking into account the RTT of the connection. A key factor contributing to a flow's end-to-end delay is the packet sojourn time within network queues, which tends to increase in congested conditions. EECN effectively minimizes this packet sojourn time, thereby reducing overall delays without compromising throughput or causing underutilization of network resources. Fig. 12a showcases the sojourn time of packets in different schemes, where EECN's performance is notably superior, leading to reduced end-to-end delays. Ensuring fair bandwidth allocation among concurrent flows is a vital aspect of any congestion control mechanism. Our proposed approach successfully ensures fairness among competing flows that utilize shared network bottleneck resources. The fair bandwidth allocation among competing flows is substantiated by the data presented in Fig. 12b. The figure shows how the 100 Mbps bottleneck link is fairly divided between the three ECN, VCP, and EECN flows, demonstrating efficient resource sharing of the proposed mechanism. Moreover, since EECN's design leverages the existing ECN bits in the TCP/IP header without adding extra signaling, this allows it to maintain performance under heavy traffic conditions, avoiding the overhead or long-term stability challenges that might arise from more complex mechanisms.

# 7. Conclusion

This work introduces the Enhanced Explicit Congestion Notification (EECN) as a promising solution to address the escalating challenges posed by network congestion. By modifying traditional ECN semantics, EECN not only notifies congestion but also provides crucial information about its severity. Notably, our proposed algorithm extends its advantages to short-lived network flows, a crucial consideration often overlooked in congestion control designs. These short-lived network flows account for most of the Internet traffic and are particularly prevalent in IoT environments. The congestion control algorithm presented, leveraging multilevel feedback from the network via EECN, exhibits significant improvements. It efficiently mitigates packet drops compared to conventional approaches, including ECN, VCP, and TCP New Reno without ECN by 70%, 95%, and 40% respectively, without compromising the throughput. It also reduces the average end-to-end delay by 19%, and minimizes the RTT of the short-flows compared to other schemes. The reduction in flow completion time by 61%, resulting in high throughput for short-lived network flows further underscores the effectiveness of our EECN-based multilevel congestion feedback solution in enhancing network performance and reliability.

## Acknowledgements

This work was supported by the Institute of Information & Communications Technology Planning & Evaluation (IITP) grant funded by the Korea government (MSIT) [2021-0-00715, Development of End-to-End Ultra-high Precision Network Technologies.]